# Fabry-Perot temperature dependence and surface-mounted optical cavities

Richard W. Fox

National Institute of Standards and Technology, 325 Broadway, Boulder, CO USA 80305

## ABSTRACT

Factors that contribute to the temperature dependence of a resonant frequency in a low-expansion optical cavity are discussed, including deformation at the cavity ends due to different coefficients of thermal expansion (CTE) of the spacer, optically-contacted mirror substrate and coating. A model of the temperature dependence is presented that incorporates finite-element-analysis of the cavity ends. A measurement of frequency versus temperature of a cavity mode is used along with the model to deduce a spacer's CTE versus temperature profile. The measured profile correlates very well with a separate experiment utilizing a temporary surface-mounted Fabry-Perot cavity fabricated on the outside of the spacer with hydroxy-catalysis bonding.

**Keywords:** Fabry-Perot, resonator, hydroxy-catalysis bonding

## 1. INTRODUCTION

Passive optical cavities are used for many applications such as optical spectrum analyzers, spectral filters and for laser stabilization. In terms of applications requiring stability, using a cavity as a reference for a high quality stabilized-laser system is perhaps the most demanding. Extremely narrow spectral linewidth atomic resonances (e.g. Hz level) that are the basis for modern atomic clocks must be probed with beams that are even narrower to take advantage of the potential stability. These spectrally pure lasers are realized by Pound-Drever-Hall (PDH) frequency locking[1] to a Fabry-Perot reference cavity in a vacuum chamber. The reference cavities have improved markedly over the years with the need for higher stability. This paper begins with a brief overview of reference cavity technology and identifies the temperature dependence of a cavity's resonances as a performance issue that generally could be improved. A related problem is that while the CTE of commercial low-expansion glass used in cavities is relatively small, the uncertainty in the CTE is too large to implement temperature independent designs at specified temperatures.

Very stable Fabry-Perot cavities are conventionally fabricated by optically contacting dielectric-coated mirrors with a flat and polished annular mounting surface onto the polished ends of a glass spacer. Exceptionally low-loss mirrors made from ion-beam-sputtered (IBS) films of $Ta_2O_5$, $SiO_2$ and other materials have made possible cavity finesses exceeding $10^5$. Thus cavities with lengths on the order of 10 cm can exhibit line-widths less than 10 kHz. A shot-noise-limited PDH locking system can tightly lock the frequency of a laser to a small fraction of the cavity resonance, providing a laser spectrum usually limited by vibration-induced length changes of the cavity, even with the cavities on isolation platforms.[2] In recent years much progress has been made in reducing vibration-induced length changes by ingenious mounting approaches that effectively cause the mirrors to move in the same direction when accelerations are applied.[3,4,5] Reductions of the resonant frequency response to vertical and horizontal vibrations by a factor of 1000 and 30 respectively have been demonstrated in a horizontally-mounted cavity.[6]

Temperature fluctuations also affect a reference cavity's fractional frequency stability, as nominally $\Delta v/v = -\Delta L/L$. To reduce the fluctuations, blackbody shielding surrounding the cavity, and drift subtraction with an acousto-optic modulator are often employed. Cavity construction with a ULE or TSG glass[7], or glass-ceramic spacer and optically contacted mirrors results in a relatively small temperature dependent fractional frequency change ($\Delta v/v$) near room temperature, typically within the range of a few parts in $10^8$ per K. Turning our attention to ULE or TSG, for a given sample near room temperature the CTE (or $\alpha(T)$) averaged along a length exhibits a positive and somewhat linear slope with temperature, crossing CTE = 0 at some temperature $T_s$ that varies from sample to sample. Consequently, versus temperature, the length of the glass spacer undergoes a minimum at $T_s$. A measurement of an optical mode's resonance-frequency versus temperature will show a near parabolic response with a frequency maximum at the so-called turning point (although not necessarily at $T_s$ due to end effects as will be discussed later). In several works it has been noted that the temperature independent point occurred at a significantly different temperature than expected from the spacer glass specification.[8,9,10] This impacts the thermal design of the vacuum system to minimize frequency drift. Clearly it would be useful to specify an optical cavity's turning point with at least modest accuracy in advance of it's installation in a vacuum chamber. This will be especially important for smaller portable systems that are constrained by power, weight and space.

The main focus of this paper is to report progress toward the ability to fabricate reference cavities with the first order temperature independence point known by design. First, a model of the resonant frequency versus temperature is presented that includes temperature-change-induced deformation at the cavity ends and other factors that cause temperature dependence. Structural finite-element analysis is incorporated into the model in an analytical fashion. Secondly, even with a valid model in hand we still need a means to obtain low expansion glass with a known $\alpha(T)$ characteristic. The tolerance of $\alpha(T)$ that commercially available glasses exhibit is simply too wide. A means to test potential glass spacer pieces by building temporary surface-mounted cavities on fine ground glass is proposed and demonstrated.

This introduction would be incomplete without noting that even with a cavity temperature-stabilized at the turning point, the frequency fluctuations are still temperature dependent for several reasons. There are temperature dependent fluctuations of the cavity length attributed to Brownian motion.[11] Secondly, laser power that is absorbed in the coatings and substrates causes length changes through the material CTE. This dynamic effect has been investigated both theoretically[12] and experimentally[13].

## 1.1 Previous literature

As will be discussed below, mirror substrates constrained by an optical contact to a spacer will cause a temperature dependent deformation of the cavity ends due to differential expansion between the spacer, substrate and coating materials. The differential expansion of a specially shaped fused silica mirror on a sapphire spacer has been analyzed and shown to move the temperature independent point of the cavity, allowing the turning point to be tuned by selecting the mirror parameters.[14] And the possibility of differential expansion between the substrates and spacer influencing the cavity length has been mentioned by many authors previously (e.g., ref 10). However, to this author's knowledge a general treatment applicable to all mirror and spacer geometries that allows end deformation to be included in the cavity's temperature dependence has not been discussed.

## 2. CAVITY TEMPERATURE DEPENDENCE

The description of the temperature dependence of the resonance frequency is developed by starting with the phase accumulated in one round-trip of a cavity of length $l$ and equating this to an integer multiple of $2\pi$:

$$\frac{2\pi 2l}{\lambda} - 2\varphi - \psi = 2\pi m. \tag{1}$$

The length $l$ is defined as the physical distance between the dielectric mirror surfaces, $\lambda$ is the wavelength in the medium, and $\varphi$ is the phase-shift upon reflection from the coatings (both mirrors assumed to be the same). $\psi$ is the phase difference between the plane wave propagation given in the first term and the actual Gaussian beam propagation. Siegman[15] gives a close approximation of $\psi$ in terms of the cavity geometry,

$$\psi = 2(1 + p + q)\cos^{-1}(1 - l/R), \tag{2}$$

where $p$ and $q$ are transverse mode indices and $R$ is the radius of curvature. Solving Eq.(1) for the frequency of the cavity resonance in vacuum one finds that

$$v = \frac{c}{2l}(m + \frac{2\varphi + \psi}{2\pi}). \tag{3}$$

Taking the derivative of Eq. (3) with respect to temperature for $\Delta l \ll l$ results in

$$\nu(T) = \nu(T_o)(1 - \frac{\frac{\Delta l}{\Delta T}\Delta T}{l}) + \frac{c}{2l}(\frac{2\frac{\Delta \varphi}{\Delta T} + \frac{\Delta \psi}{\Delta T}}{2\pi})\Delta T, \qquad (4)$$

where $\Delta T = T - T_o$. Eq.(4) gives the expected resonance frequency at some temperature $T$ as a function of the resonance frequency at an arbitrary temperature $T_o$. The second term is dominated by the change in the coating phase shift, as $\Delta \varphi/\Delta T \gg \Delta \psi/\Delta T$. The change of $\varphi$ with temperature can be estimated from the slope of the spectral phase shift upon reflection and the temperature response of the high-reflectivity mirror's band-pass. The mirrors used in the experiments to be described were high-reflectivity visible band (centered at 578 nm) IBS coatings on low expansion substrates. At the measurement wavelength of 633 nm, $\Delta \varphi/\Delta T \cong -0.2$ mr/K. The corresponding frequency shift depends on the free-spectral-range (FSR), and the cavities described are about 10 cm long. The frequency shift contribution from the last term (approximately -95 kHz/K) is small in comparison to the measured frequency shift that will be shown below.

As the temperature changes uniformly, the cavity's physical length will change due to the spacer expansion or contraction but there are other factors that cause the mirror spacing to change. The elastic modulus and Poisson's ratio of fused silica (the primary constituent of ULE and TSG) are both slightly temperature dependent (~0.02%/K). This may translate to a temperature dependent change in the cavity length under gravity depending on how the cavity is supported. And a gradient in the spacer CTE that is transverse to the optical axis would cause the spacer to bend with temperature. As mentioned previously, structural deformation at each end of the cavity due to the different CTE of the spacer, mirror substrate, and coating is also a factor that causes the mirror separation to vary with temperature. The mirror substrate may be a different material than the spacer purposely for low-noise considerations.[11] But use of the same type of glass for substrate and spacer does not preclude distortion at the ends. Indeed, with a ULE substrate and spacer at the ideal zero expansion temperature $T_s$, finite element analysis (of the structure presented below) reveals an on-axis inward deformation that is more than twice as large as the linear thermal expansion of the 5 μm thick coating. However, it is much more likely that a ULE mirror substrate will actually have a different coefficient of expansion than the end of a spacer made from ULE. Often this arises because the mirror substrate material is acquired from a different source than the spacer, allowing for the distinct if not probable possibility that the two pieces of glass are from different boules entirely. Even if the substrate glass is cut from the part of the boule adjoining the spacer glass, the spatial variation of $\alpha(T)$ in ULE can be several $(10^{-9})$/K per cm.[16]

In general then, temperature dependent deformation of the cavity ends contributes to $\Delta l$. This is treated here with finite element analysis (FEA) under the assumption that optical contacting constrains the substrate to the spacer and does not slip as temperature-induced stress builds up. No frequency discontinuities that would indicate slipping were observed over the limited temperature range measured and reported on here. The assumptions relied on here may not be valid for every cavity or for wide temperature excursions.

The physical length change, $\Delta l$, may be written as the sum of the change of the spacer's length and also contributions from the ends, $\Delta l = \Delta l_S + 2\Delta l_E$. The factor of 2 is because a symmetrical cavity is assumed. The sign of $\Delta l_E$ is defined such that a positive change at either end makes the cavity longer. We explicitly include these contributions by rewriting Eq.(4) as

$$\nu(T) = \nu(T_o)(1 - \frac{(\frac{\Delta l_S}{\Delta T})\Delta T}{l} - 2\frac{(\frac{\Delta l_E}{\Delta T})\Delta T}{l}) + \frac{c}{2l}(\frac{2\frac{\Delta \varphi}{\Delta T}\Delta T}{2\pi}). \qquad (5)$$

The part of the last term concerning the diffraction phase shift has been dropped as it is not significant. The length temperature dependences $\Delta l_S/\Delta T$ and $\Delta l_E/\Delta T$ are functions of temperature because the glass CTE is temperature dependent. Eq.(5) can be used to predict resonance temperature dependence from an arbitrary $\nu(T_o)$ if the variables can be expressed as a function of temperature. Importantly, if the variables can be written analytically, then Eq.(5) can be used by a least-squares routine to compare the model with measured data. For instance, starting from a measured point $\nu(T_o)$, a routine could adjust parameters in the model to match $\nu(T)$ to the data.

Writing $\Delta l_S$ as a function of the initial and final temperature is straightforward as the CTE of spacer glasses can be represented by a polynomial, for instance $\alpha(T) = K_o + K_1 T + K_2 T^2$. The fractional change in length is given by integrating the CTE over the temperature interval,

$$\Delta l_S = l_S \int_{T_o}^{T} \alpha(T) dT . \tag{6}$$

Determining $\Delta l_E / \Delta T$ as a function of temperature can be accomplished by FEA of the end structure to the extent that the material parameters (CTE, Elastic modulus, Poisson's ratio) of the spacer, substrate and coating are known. However the crux of the matter is that in many instances the CTE of either the spacer or the mirror substrates or both is not known with sufficient precision to make any useful prediction. Examples are the tolerance of the average CTE value of ULE, ±30 ppb/K, or that of Zerodur[7], ±20 ppb/K. But, as noted already, if $\Delta l_E(T)$ can be expressed in an analytical form then Eq.(5) offers a means to analyze experimental data to measure the CTE of components. That is achieved here by first using FEA to find an analytical approximation of the z-axis deformation as a function of the CTE values of the spacer and substrate, independent of temperature. Then, given the task of determining the end deformation that arises from $T_o$ to a temperature $T$, this approximation can be used in conjunction with the CTE equations of the materials involved to provide an analytical form of Eq.(5).

As an example we analyzed optical frequency versus temperature data from a cavity fabricated with a TSG spacer and contacted high-finesse fused silica mirrors (Fig.1). The CTE of the TSG bar was unknown other than the material specification of a mean CTE in the range of ±100 ppb/K over the temperature interval from 5 to 35 C. The measurement was performed by recording the beat-note between an Iodine-stabilized HeNe laser and an extended cavity diode laser that was frequency locked to a longitudinal mode of the cavity. The cavity was in a thick-walled blackbody vessel inside a vacuum chamber (P < 1 Pa) as the outside temperature of the chamber was very slowly ramped. The cavity was supported on a soft rubber o-ring in a v-block to avoid imparting the expansion of the mounting to the cavity. A parabolic fit to the Fig. 1 data has a turning point at -22.2 C and a quadratic coefficient of -346 kHz/K$^2$.

The physical geometry is modeled by supporting the TSG bar at one end and calculating the z-axis deformation of the center of the opposing coated mirror for a selected set of spacer and mirror substrate CTE values ($\alpha_S$ and $\alpha_M$ respectively) given in Table I. The deformation is calculated for a +1 degree $\Delta T$ and the units are nm/K. This is followed by accounting for the spacer expansion or contraction by suppressing the opposing mirror and calculating the z-axis position of the end of the spacer, again in nm/K. The difference between the two calculations is taken as the end deformation corresponding to this physical structure and coating with those particular values of $\alpha_S$ and $\alpha_M$. The HR quarter-wave stack (~60 layers) is modeled as a uniform 5 μm thick layer on the central 18 mm of the substrate diameter. The coating parameters are taken as $\alpha_c = 1.8 \times 10^{-6}$ K$^{-1}$, $E = 106$ GPa, $\sigma = .21$ and the density = 8200 kg/m$^3$ following work on similar coatings,[17] although these assumptions are a possible source of error as the physical constants of thin-films are typically less well known than the bulk material. Care was taken to insure an adequate FEA mesh around the contacted regions of the components.

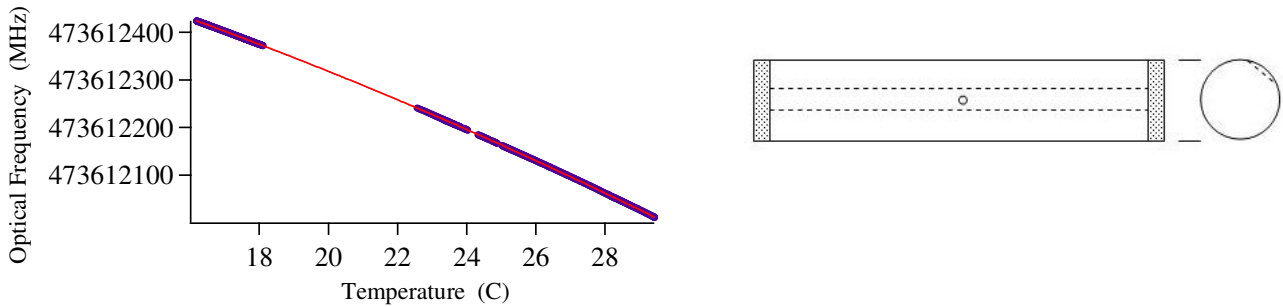

Fig. 1 (Left) The thick line is the optical frequency versus temperature data for a mode of the TSG glass cavity, the thin line is a fit to the model described in the text. (Right) The cavity consisted of a 99 mm long TSG spacer, 25.4 mm diameter, with polished ends, a 6.5 mm dia center bore and an air relief hole. Fused silica mirrors with an outer flat annulus (25.4 mm dia, 5.75 mm thick, 50 cm radius) were contacted to the ends (contacted at radius > 9.5 mm). The spacer had an 8 mm wide flat ground along the entire length as depicted by the dashed line in the end view for the placement of the surface mounted mirrors described in section 3.

| Spacer CTE (ppb/K) | Substrate CTE (ppb/K) | $\Delta l_E/\Delta T$ From FEA (nm/K) | Eq.(7) error % |
|---|---|---|---|
| +30 | 460 | 2.6549 | 0.039 |
| +30 | 480 | 2.7823 | -0.012 |
| +30 | 500 | 2.9077 | .009 |
| +30 | 520 | 3.0346 | -0.021 |
| +10 | 460 | 2.7766 | -0.074 |
| +10 | 480 | 2.9022 | -0.056 |
| +10 | 500 | 3.0281 | -0.051 |
| +10 | 520 | 3.1545 | -0.061 |
| -10 | 460 | 2.8898 | 0.117 |
| -10 | 480 | 3.0161 | 0.102 |
| -10 | 500 | 3.1419 | 0.105 |
| -10 | 520 | 3.2676 | 0.110 |
| -30 | 460 | 3.0138 | -0.068 |
| -30 | 480 | 3.1396 | -0.058 |
| -30 | 500 | 3.2653 | -0.046 |
| -30 | 520 | 3.3909 | -0.033 |

Table 1. Finite element analysis calculations of the end deformation over a +1 K temperature change are given in the third column. In each case the spacer and mirror substrate CTE are held constant at the values given in the first two columns. The data in the third column can be approximated by Eq.(7). The fourth column gives the error of this approximation.

With the values of $\Delta l_E/\Delta T$ calculated by FEA in hand we can approximate them with a simple analytical form

$$\frac{\Delta l_E}{\Delta T} = A\alpha_M + B\alpha_S + C . \quad (7)$$

In eq.(7) the units of the CTE terms are ppb/K, the resulting deformation $\Delta l_E/\Delta T$ is in nm/K, and the coefficients $A$, $B$, and $C$ are found by a least-squares fit to the calculated points to be $A = 0.0063$ nm, $B = -0.0059$ nm, and $C = -0.064$ nm/K. Indeed as the last column of the table shows, at least over the limited range of CTE values in the table Eq.(7) is a very good approximation to the end deformation calculated by FEA. These coefficients are unique to the physical structure and coating analyzed.

The CTE terms $\alpha_M$ and $\alpha_S$ in Eq.(7) can be represented by temperature dependent polynomials, thus providing the analytical representation of the end deformation. The $\Delta l_E$ term in Eq.(5) can be calculated for an arbitrary temperature by utilizing eq.(7) and integrating from $T_o$ to $T$.

The fused silica polynomial, $\alpha_M$, is relatively well known and so the unknown TSG spacer polynomial coefficients ($K_o$, $K_1$ and $K_2$) can be determined by fitting Eq.(5) to the data shown in Fig. 1. The uncertainty in this process is due primarily to the uncertainty of the fused silica CTE. Two sets of best-fit spacer coefficients were determined (Fig. 2) since published values of the fused silica CTE near room temperature differ slightly, perhaps due to SiO$_2$ purity issues.[18] The best-fit result is that the TSG spacer zero-crossing point $T_S$ is in the range of 16 to 20 C, limited by the uncertainty in the CTE of the fused silica substrate. Thus the model indicates that the spacer zero crossing $T_S$ is approximately 38 to 42 K higher than the optical cavity's turning point of -22 C.

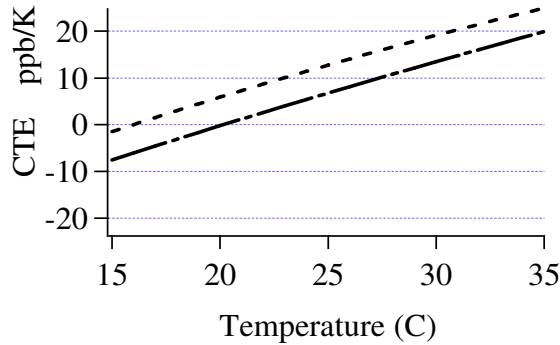

Fig. 2. Two estimates of the TSG spacer CTE from optical frequency measurements after accounting for end deformation. The short dashed line is with fused silica $\alpha_M(T)$ per ref [18], and the long dashed line is with $\alpha_M(T)$ per the discontinued NIST Standard Reference Material 739. The NIST SRM CTE values exceed than those of ref [18] by about 8 %.

## 3. SURFACE MOUNT CAVITIES AND TEMPORARY FABRY-PEROTS

For some time at NIST we have been experimenting with building optical cavities on flat polished surfaces as shown in Fig. 3. The mirrors have been cut off and polished to provide a flat mounting surface that is parallel to the optical axis to within ±10 arc minutes. The mounting surfaces have been specified flat to $\lambda/8$ with a 60/40 surface finish. With two mirrors placed on the surface of the flat spacer it is possible to manually orient the mirror axis with the horizontal degree of freedom to form a cavity, and align an incoming beam such that cavity modes are evident on the transmitted beam. This construction has been investigated for use as a wavelength reference for stage interferometers, because it can provide a mechanically stable cavity that is also open to the free flow of air.[19] Although the specified flatness and surface finish are sufficient for optical contacting, the mirrors are not attached by conventional contact since alignment would be difficult. Instead, hydroxy-catalysis bonding using a potassium hydroxide (KOH) solution is employed.[20] This provides an adhesive-free permanent bond that also allows the mirror positions to be adjusted (at least for several minutes) after they have been set on the surface.

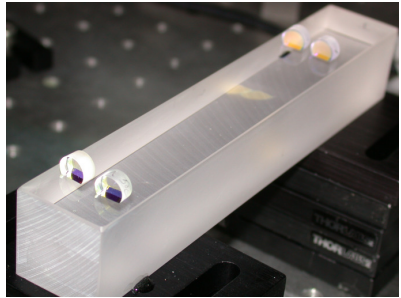

Fig. 3  (Left) Four 50 cm radius (7.75 mm dia) mirrors arranged in a surface-mounted bow-tie ring cavity fabricated on a 15 cm long ULE block. The mirrors are aligned manually in the horizontal direction by gentle forces applied with a hand-held probe.

Cleanliness and a particle-free environment are integral to this process as in conventional optical-contacting. Bonding is accomplished one mirror at a time by applying approximately 0.5 μL of KOH/$H_2O$ solution (ph = 12.5) to the interface after cleaning, then re-aligning the mirror before the bonding sets. As discussed in reference [20] the bond curing is a de-hydration process. The permanently attached mirrors appear to have no internal reflection at the interface if the bond completely covers the interface once cured. However partial bonding over the interface area is possible. This is easily detected, for if the bonding surfaces are non-ideal in terms of flatness or particle contamination, the non-bonded portion of the interface will reflect light.

Our concern here is the application of the surface-mount approach to fabricating a Fabry-Perot test cavity on the non-polished (fine ground) glass surface of the TSG spacer. Specifically, a measurement of a test cavity built on the TSG spacer surface could corroborate the spacer's longitudinally averaged CTE measured in section 2. Furthermore, a demonstration of such a metrology capability with temporary mirrors is important because it offers another means to measure the CTE of potential spacer and mirror glass prior to expensive machining.

Uncertainty in the measurement would be reduced if the mirrors were attached to the ground glass such that $\Delta l_E$ of Eq.(5) is negligible. In other words, mounted in such a manner that the distance between the coated surfaces depends only on the spacer. As with mirrors contacted on the spacer ends, that is generally not true of the surface mounted mirrors bonded to a polished surface. FEA modeling of a fully-bonded surface-mounted substrate with a different CTE than the spacer indicates that at the position of the cavity mode in the center of the mirror, several mm away from the constrained interface, the substrate expands freely with temperature. Therefore unless the ULE mirror substrate CTE is well known this introduces an uncertainty in the desired measurement of the spacer CTE. The same principle applies with a surface mounted mirror attached via hydroxy-catalysis bonding to a fine ground surface. One approach to minimize the uncertainty is to bond the test mirror near the coated surface only, allowing the rear portion to freely expand. For this purpose a saw cut was introduced as shown in Fig. 4, and a minimal amount of KOH liquid applied (~0.1 μL). The solution wicked in to the thin front portion when applied to the bottom of the coated surface as the mirror was properly positioned on the fine ground surface of the TSG. After 30 minutes the bond was tested by holding the spacer upside down to check that the mirrors were in fact attached. The TSG spacer with the surface-mounted cavity was then placed in the chamber as in the previous test and the frequency data recorded. As shown on the right in Fig. 4, the temporary cavity has a turning point at 17.5 C, fully consistent with the previous conclusion of the TSG spacer CTE based on the model. After the measurement was complete, the spacer was removed and an inspection determined that the surface-mounted mirrors were still bonded. (Our limited experience to date with KOH bonding on non-polished surfaces is that the bond does not always set properly). The substrates were removed with minimal force and appeared unharmed, as did the spacer surface.

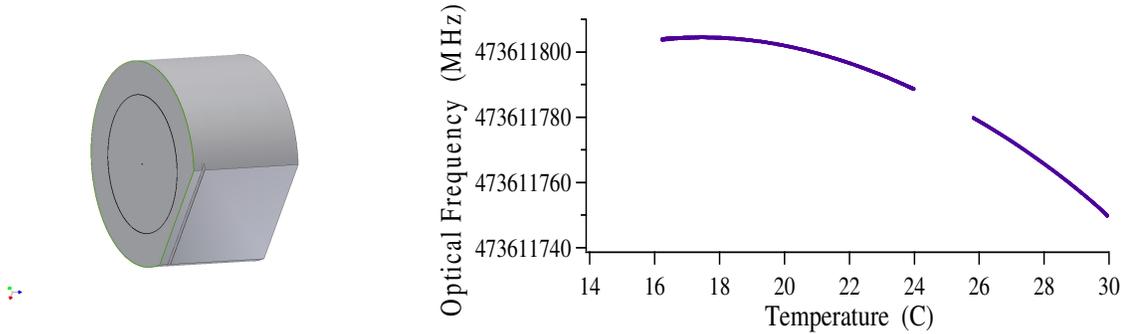

Fig 4. (Left)   A drawing of a saw cut near the front of a surface mounted mirror for the purpose of limiting the bonding to the edge nearest the coating. (Right)   Optical frequency versus temperature of a temporary cavity bonded to the TSG spacer using two such saw-cut mirrors. The data points were spaced by 5 minute intervals over a three day period except when the laser unlocked. The turning point is approximately 17.5 C, in very good agreement with Fig. 2.

## 4. CONCLUSION

A method to incorporate finite element analysis calculations of temperature-induced deformation at the ends of optical cavities has been presented. This may prove to be a useful tool in the measurement of low-expansion material constants, as similar optical cavity frequency measurements have been used for CTE calibration. The analysis may also be a useful tool for the design of optical cavities with tightly specified turning point temperatures. The method was used to measure the CTE versus temperature profile of a TSG glass spacer. Separately, a temporary Fabry-Perot cavity was fabricated on the side of the TSG spacer enabling a second measurement of the spacer CTE, in very good agreement with the first. Such temporary cavities could be used to tighten the CTE tolerance of prospective pieces of cavity glass prior to machining. Together, the analysis method and tighter tolerances on low-expansion CTE materials provide a path to fabricating optical cavities with temperature independence at specified temperatures.